# Towards Compact and Frequency-Tunable Antenna Solutions for MIMO Transmission with a Single RF Chain

Mohsen Yousefbeiki, *Student Member, IEEE,* and Julien Perruisseau-Carrier, *Senior Member, IEEE*

*Abstract*—Recently, a technique called beam-space MIMO has been demonstrated as an effective approach for transmitting multiple signals while using a single RF-chain. In this work, we present novel design considerations and a compact antenna solution to stimulate the deployment of beam-space MIMO in future wireless applications. Targeting integration in small wireless devices, the novel antenna is made of a single integrated radiator rather than an array of physically-separated dipoles. It also drastically simplifies the implementation of variable loads and DC bias circuits for BPSK modulated signals, and does not require any external reconfigurable matching circuit. Finally, we show that this antenna system could be reconfigured by dynamic adjustment of terminating loads to preserve its beam-space multiplexing capabilities over a 1:2 tuning range, thereby promoting the convergence of MIMO and dynamic spectrum allocation via reduced-complexity hardware. A prototype achieving single-RF-chain multiplexing at a fixed frequency is designed and measured, showing excellent agreement between simulations and measurements.

*Index Terms*—Beam-space MIMO, cognitive radio, dynamic frequency allocation, MIMO antenna, reconfigurable antenna, reduced complexity.

## I. INTRODUCTION

MULTIPLE-INPUT multiple-output (MIMO) techniques are now making their way to real applications as they allow drastic improvements in the reliability and/or transmission rate of wireless communication systems [1]. This is also motivated by the ever-increasing number of wireless devices deployed (i.e. higher interference), the demand for higher data-rates, and scarce spectrum and energy resources. However, there are some major challenges towards the implementation of non-distributed MIMO concepts in various wireless communication systems, especially in small and low-cost terminals. First, in general multiple antennas should be placed far enough away from each other to reduce mutual coupling and obtain almost decorrelated MIMO channels. Even more important in practice, MIMO traditionally demands multiple radio frequency (RF) chains, resulting in greater complexity, larger power consumption, and higher hardware cost. While the first problem can somehow be solved with different techniques, such as decoupling networks [2], [3], and pattern orthogonality approaches [4], [5], the need for multiple RF-chains remains for implementing arbitrary spatial multiplexing and diversity MIMO schemes. Different design alternatives have been proposed to overcome the second challenge: antenna selection [6], analogue antenna combining [7], time-division multiplexing [8], and code-modulated path-sharing [9]. However, these techniques do not allow multiplexing signals using a single-RF chain. Finally, [10] proposed a low-complexity MIMO transmission technique named spatial modulation (SM), which results in a logarithmic increase of the data rate with the number of transmitting antennas. Since it requires multiple antennas at the transmitter, it is more suited to base stations than small portable devices.

Recently, a novel concept has been proposed for transmitting multiple data streams while using a single RF-chain and a switched parasitic array [11]–[14]. This technique, referred to as beam-space MIMO, was shown to have the potential of simultaneously addressing both the size and multiple RF-chains challenges of classic MIMO techniques. Fig. 1 is symbolic representation of a beam-space MIMO transmitter. Instead of driving different symbol streams to different active antenna elements as in traditional MIMO transmission, symbols are mapped directly onto an orthogonal set of basis functions in the beam-space domain of a single active antenna. This principle is implemented in practice by designing an antenna system whose instantaneous pattern, $G_T$, can be decomposed at any instant of time into an *orthogonal basis* formed by $B_1$ and $B_2$, such that $G_T = s_1B_1+s_2B_2$, $s_1$ and $s_2$ taking any value required by the considered modulation. The authors in [11] introduced the general concept of beam-space MIMO on a theoretical level. However the design was based on simple analytical description of ideal omni-directional elementary radiators and only in a 2D plane, which is not realistic for practical applications. The first successful experimental demonstration of the concept was then reported in [14], where a symmetric switched parasitic array of three printed dipoles was employed for multiplexing two binary-

Manuscript received September 24, 2012; revised January 29, 2013 and April 26, 2013; accepted June 3, 2013. This work was supported by the Swiss National Science Foundation (SNSF) under grant n°133583.

The authors are with the group for Adaptive MicroNanoWave Systems, LEMA/Nanolab, Ecole Polytechnique Fédérale de Lausanne (EPFL), CH-1015 Lausanne, Switzerland (e-mail: mohsen.yousefbeiki@epfl.ch, julien.perruisseau-carrier@epfl.ch).

Digital Object Identifier 10.1109/TAP.2013.2267197

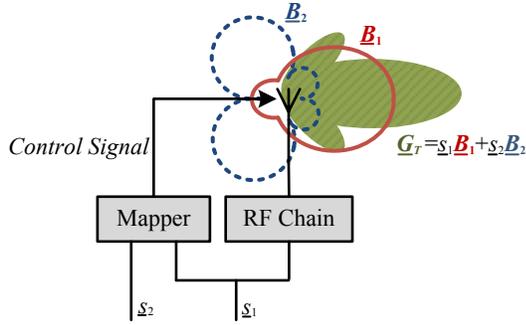

Fig. 1. Symbolic representation of the beam-space MIMO concept. Each complex symbol stream $s_1$ and $s_2$ is assigned to a virtual antenna of complex basis patterns $B_1$ and $B_2$. The orthogonality of the basis patterns ensures decorrelation of the MIMO channel elements in a rich-scattering environment, and allows decoding the multiplexed symbols at the receiver.

phase shift keying (BPSK) data streams over-the-air. The central dipole element was fed with a RF signal modulated by the first baseband data stream, while the terminating loads of other two parasitic elements were controlled by a baseband signal containing information related to the ratio of the second and the first data streams $s_2/s_1$.

However, the parasitic array antenna system in [14] is also unsuitable to small wireless portable devices, mainly due to its size and the dependence of its properties to the immediate environment. Moreover, it requires the implementation of a complex DC biasing network to decouple driving signals from RF load circuits and radiating structures. In this paper, we present an antenna system solution for beam-space MIMO that can be effectively incorporated in a compact, low-cost portable terminal. The solution also enables an extremely simple biasing scheme as explained next. Instead of a set of dipole and monopole radiators, a compact multi-port integrated antenna structure is used. Moreover, for low-order modulations, this approach allows to drastically simplify the reconfigurable loads. Finally, we show that it is possible to use the designed antenna to achieve the multiplexing gain at a dynamically-controllable frequency. This result is very promising for the future convergence of MIMO and cognitive frequency allocation concepts via low-cost reduced-complexity transceivers.

The rest of the paper is organized as follows. Section II first introduces the proposed compact antenna solution for beam-space MIMO. Then, in order to experimentally validate the solution, we present the design of a compact antenna prototype, integrated to a typical portable platform, and discuss the simulated and measured return loss and radiation patterns. In Section III, we investigate frequency reconfigurability of the antenna system example to enable dynamic frequency allocation in beam-space MIMO transmission. Finally, Section IV concludes the paper.

## II. COMPACT ANTENNA SOLUTION FOR BEAM-SPACE MIMO

### A. Single and Compact Radiator Design

Consider a reconfigurable antenna system which has a single active input port and allows the formation of two radiation patterns $G_1$ and $G_2$ that are *mirror* images of each other, for instance $G_2(\vartheta,\varphi) = G_1(\vartheta,\pi-\varphi)$ for a plane of symmetry at $\varphi=-90°\text{-}90°$. It can be shown that the angular functions defined as

$$B_1(\vartheta,\varphi) := \frac{1}{\sqrt{2}}\left[G_2(\vartheta,\varphi) + G_1(\vartheta,\varphi)\right]$$
$$B_2(\vartheta,\varphi) := \frac{1}{\sqrt{2}}\left[G_2(\vartheta,\varphi) - G_1(\vartheta,\varphi)\right] \quad (1)$$

form an orthogonal set of basis functions [14], namely $\int_\varphi \int_\vartheta B_1(\vartheta,\varphi) \cdot B_2^*(\vartheta,\varphi) \sin\vartheta \, d\vartheta \, d\varphi = 0$. Let us emphasize that instantaneous and basis patterns in (1), namely $G_1$, $G_2$, $B_1$, and $B_2$, are the *complex* vector radiation patterns.

When the active port is fed by the first signal, it is possible to simultaneously transmit two BPSK signals $s_1$ and $s_2$ which are independently mapped onto each basis pattern such that (see [14]):

$$G_T(\vartheta,\varphi) = \frac{1}{\sqrt{2}}\left[s_1 B_1(\vartheta,\varphi) + s_2 B_2(\vartheta,\varphi)\right]. \quad (2)$$

However, the orthogonality of the basis does not ensure impedance matching at the active port nor an equal power distribution across the basis functions. Therefore, a performance optimization must then be carried out, as further discussed below.

As aforementioned this technique was implemented in [14] using a three-element switched parasitic array of printed dipoles. However, this type of antenna is too large to be utilized in modern compact, low-cost portable devices. Moreover, the requirement of utilizing several disconnected radiators (active and passives ones) is inconvenient not only in terms of size but also in terms of fabrication or sensitivity to the influence of the user on the antenna performance. On the other hand, nowadays tough requirements of modern communication devices demand that a built-in antenna must be treated as a component of the whole platform rather than an isolated element within the design process.

Therefore, the first contribution proposed in this work is an antenna solution for beam-space MIMO that can be effectively incorporated in a compact, low-cost portable terminal. Rather than a three-element structure in which each element is regarded as a distinct and physically-separate radiator, a symmetric three-port antenna structure is used, as symbolized in Fig. 2a, integrated into a device platform. The central port (port 0) is the active input, while the other two are passive and loaded with pure imaginary loads as the real part of a complex load would degrade the efficiency of the antenna system. Obviously, by permuting the reactive loads, $X_1$ and $X_2$, the antenna system is capable to create two mirrored instantaneous patterns with regard to the plane of symmetry depicted in Fig. 2a, namely $G_2(\vartheta,\varphi) = G_1(\vartheta,\pi-\varphi)$ (for a plane of symmetry at $\varphi=-90°\text{-}90°$). Therefore, when the central port is fed with the first symbol stream, the antenna system can simultaneously transmit two streams and map them onto an orthogonal basis, as explained above. An example of this solution for beam-space MIMO has been designed and successfully measured, as detailed in Section II.*C*.

### B. Simplified Variable Load Implementation

Prior design of beam-space MIMO considered a given a-priori antenna structure, based on which the variable load impedance parasitic values are optimized [12], [14]. With this approach, the reactance values required for the optimum operation of the antenna system differ from those of available control elements, such as for instance p-i-n diodes. As a consequence, another step in the antenna system design procedure must be devoted to the realization of desired variable loads by a circuit embedding the control devices as well as other passive elements. The proper design and operation of this load control circuit is critical for the actual implementation of a switched antenna structure in terms of design effort, system performance, and cost [14].

In this context, here we propose to use the control elements such as p-i-n diodes directly as the variable loads in the design of a switched antenna system for beam-space multiplexing, thereby avoiding the design of the aforementioned complex variable load circuits. This is achieved by optimizing the antenna structure itself for the desired loads (e.g. the impedances corresponding the each of the two states of the p-i-n diodes as depicted in Fig. 2b), rather than the other way around as done in prior art [11]–[14]. Although this approach put more burdens on the design of the antenna structure, reducing the complexity of the load circuits effectively reduce the overall design complexity, while also increasing the certainty and consistency in the proper operation of the beam-space MIMO terminals (in the future the joint optimization of the loads and the antenna structure could prove an interesting solution as well). Such an approach is easily implemented for BPSK modulation, as shown in the following subsections. It could then be extended to higher order modulation by using a simplified load consisting for instance of multiple diodes connected in parallel.

### C. Antenna Design and Structure

In this subsection, we implement the solutions presented in the two previous subsections by designing a compact switched antenna prototype for the beam-space multiplexing of two BPSK signals in a hypothetical portable platform. The antenna prototype is designed for an uplink frequency range of 1920-1980 MHz and a downlink frequency range of 2110-2170 MHz, and realized in close proximity to a grounded 1.6-mm-thick FR4 substrate of 75 mm × 50 mm. A compact planar radiating architecture is selected, as depicted in the unfolded view of Fig. 3. According to the solution explained in detail in Section II.*A*, the antenna system consists of a single symmetric three-port radiator. The central port is matched to a 50-Ω coaxial excitation without any requirement for external matching circuit.

Thanks to the design approach described in Section II.*B*, two fast-switching silicon p-i-n diodes (Aeroflex/Metelics MPN7310A-0805-2) are used as the sole mounted components at the two passive ports. The p-i-n diodes are embedded in the antenna system in such a way that in each instant, one of the diodes is forward biased and the other is reverse biased. Therefore, by switching the states of both p-i-n diodes at the same time, the instantaneous antenna radiation pattern becomes mirrored thanks to the symmetry of the structure. As depicted in Fig. 3, there is a DC path from each p-i-n diode to the ground plane of the PCB and thereby to the outer conductor of the coaxial connector through a metalized via and a RF choke (Murata LQW15AN82N) mounted on the backside of the substrate. Therefore, the bias voltages for controlling the states of the p-i-n diodes can be applied through the connector inner conductor using a simple bias tee. In other words, the state of the antenna system is simply selected by the polarity of the DC bias source. This again allows significant complexity reduction as no DC decoupling biasing circuit must be introduced in the antenna structure.

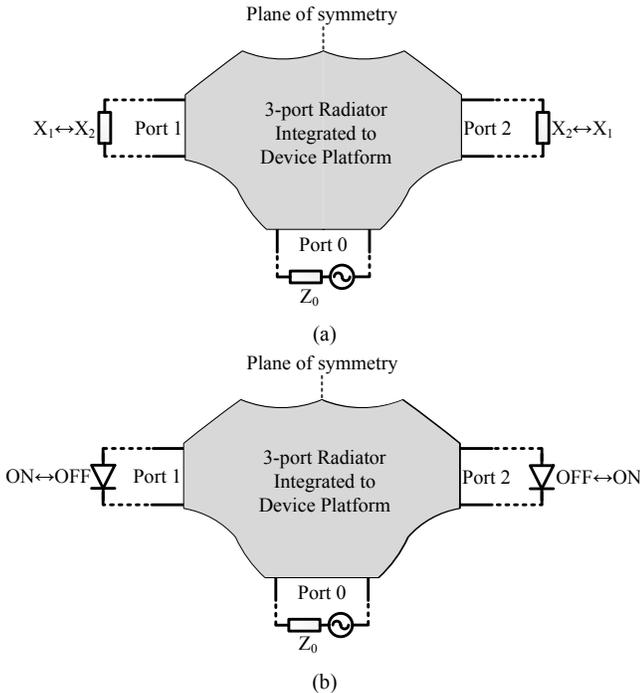

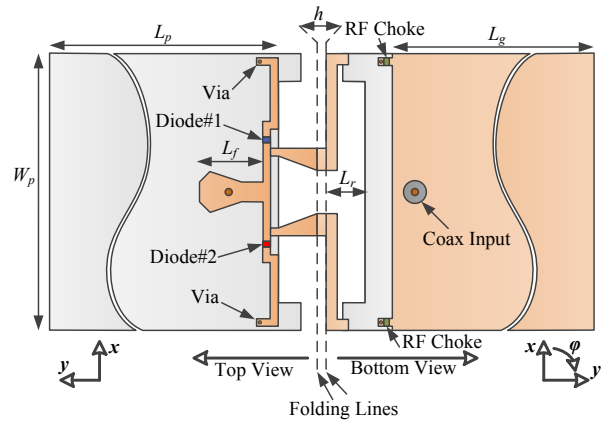

Fig. 2. Proposed antenna system solution for beam-space MIMO. (a) A built-in three-port antenna system. It is generally connected rather than physically-separate multi-element arrays. (b) Employing the p-i-n diodes directly as optimum variable loads to obviate complex load circuits and corresponding DC biasing.

Fig. 3. Unfolded schematic of antenna system prototype for beam-space multiplexing of two BPSK (Some important parameters: $W_p$ = 50 mm, $L_p$ = 80 mm, $h$ = 1.6 mm, $L_r$ = 7 mm, $L_f$ = 11.45 mm, $L_g$ = 75 mm). A part of the radiator is directly printed on both sides of ungrounded portion of the substrate, while the other part (the arms) can easily be realized by the use of beryllium-copper (BeCu) strips.

Full-wave analyses of the antenna structure were carried out using ANSYS HFSS. For a precise modeling of the embedded lumped components, i.e. the p-i-n diodes and the RF chokes, their equivalent circuit models were extracted from separate measurements using a TRL calibration technique. The impedances of the p-i-n diodes for a forward current of 19 mA and a reverse voltage of –0.6 V at the design frequency of 1.95 GHz were measured 1.9 + $j$17 Ω and 35.4 – $j$407 Ω, respectively. Then, the measured models were corrected for their insertion in the simulation according to the method explained in detail in [15].

For optimum operation of a beam-space MIMO system in open-loop scenarios with rich scattering, balanced basis patterns are desired [14]. In addition, the antenna should yield an acceptable level of the total transmit efficiency (product of radiation and matching efficiencies) at the single active port. Therefore, the optimization criterion is to minimize the power imbalance ratio between the basis patterns in the upload frequency band while simultaneously achieving a return loss better than 10 dB over both working frequency ranges. It is worth noting again that since the p-i-n diodes are here directly utilized as the sole lumped components for the pattern reconfigurability, the optimization approach should be employed for the three-port radiator itself rather than the variable loads. Accordingly, extensive full-wave simulations on different radiator parameters were performed. Then, the obtained S-parameters and radiated fields were used to find the optimal radiator parameters.

Fig. 4 illustrates the surface current density on the antenna conducting traces at the uplink and downlink center frequencies, when the upper and lower p-i-n diodes are forward and reverse biased, respectively. This is representative of the antenna operation when transmitting *any* of the BPSK signal pairs $s_1$ and $s_2$ since the radiator is symmetrical and according to Figs. 1 and 2b in all four possible combinations of BPSK signal pairs one of the p-i-n diodes is on and the other is off. It can be observed in Fig. 4 that the p-i-n diode in off state roughly acts as an open circuit, thus the lower stub (part D in Fig. 4) has a negligible effect on the antenna functionality. To illustrate the effect of the arms and the stubs and understand the corresponding antenna impedance behavior, we analyzed the antenna reflection coefficient in three different configurations: the antenna structure without the upper arm and stub (parts B and C), the antenna structure without the lower arm (part E), and the whole antenna. As depicted in Fig. 5, in the absence of the upper arm and the upper stub, the antenna shows a deep resonance in the downlink frequency band. This is well in accordance with the surface current density at 2.14 GHz shown in Fig. 4: the current is distributed mostly on the lower arm and less on the upper stub and arm. The results imply that the lower arm dimensions affect the antenna second resonant frequency, however, the presence of the upper arm and stub causes a slight decrease of the resonance level, as shown in Fig. 5. This figure also shows that the antenna structure without the lower arm yields a moderate resonance at the uplink frequency. Then, when the lower arm is included in the structure, this resonance is well tuned to the upload center frequency and its level is improved. Therefore, it can be inferred that the first resonance of the antenna structure is nearly determined by the dimensions of the upper stub and arm as well as the impedance of the p-i-n diode in on state, and adjusted by the lower arm. This is also in agreement with the current distribution shown in Fig. 4. We finally recall here that the above explanation is related to the antenna matching in each of the two symmetrical antenna states actually used for the beam-space MIMO operation. The fully-operational fabricated built-in antenna system is shown in Fig. 6.

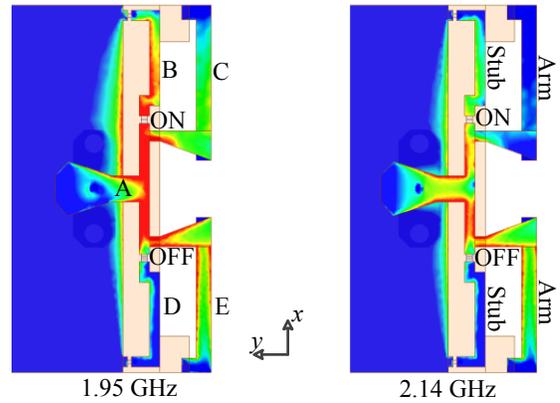

Fig. 4. Current distribution on the conductors of the antenna prototype; the upper/lower p-i-n diodes are forward/reverse biased (on/off), respectively.

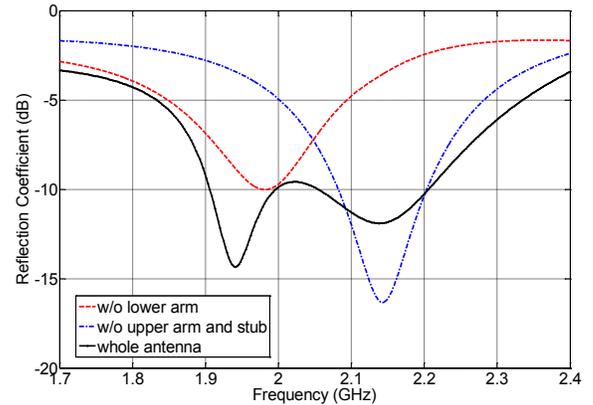

Fig. 5. Effects of the stubs and the arms on the antenna frequency-domain impedance response.

*D. Results and Discussion*

Fig. 7 shows the comparison between the measured and simulated reflection coefficients of the antenna system around the design frequency bands. The agreement between the simulations and the measurements is excellent, revealing the accuracy of the technique adopted for the modeling of the antenna system. Furthermore, the similarity between the measured results for both states confirms the symmetry of the fabricated antenna system. It is observed that the measured return loss is better than 10 dB for both system states over the upload/download bands. A measured bandwidth of 14.5% is obtained for a reference of −9 dB.

The co- and cross- polarized radiation patterns of the antenna system in the plane of the device platform are depicted in Fig. 8. Here again very good agreement is obtained between measurements and simulations. Moreover, it can be seen that the desired mirrored pattern pair is obtained when the system state is changed, as expected due to the symmetrical structure of the antenna system (with regard to the plane of symmetry, $\varphi=-90°$-$90°$). This implies that the antenna system is capable to create an orthogonal basis onto which the signals are mapped, as discussed earlier in Section II.*A*, and experimentally proved by over-the-air multiplexing and decoding in [14]. The small discrepancy between the measured and simulated cross-polarized far fields can be safely attributed to spurious radiation effects of the measurement setup (including the supporting structure, the coaxial cable and the bias tee) and antenna tolerances.

Simulations show a radiation efficiency of 76-83% and 87-89% for upload and download frequency bands, respectively, which are in good agreement with the values obtained from gain measurements. The corresponding losses can be divided into two parts: the loss introduced by the resistance of the lumped components, i.e. two p-i-n diodes and two RF chokes impedances, and the thermal loss in the dielectric substrate and metallized parts. As shown in Table I, the lumped components

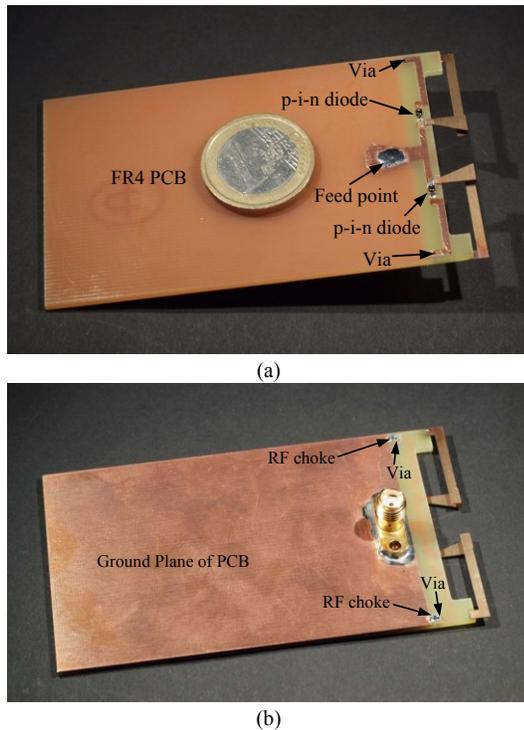

(a)

(b)

Fig. 6. Antenna system prototype for beam-space multiplexing of two BPSK signals, integrated to a hypothetical handheld device; (a) top view, (b) bottom view.

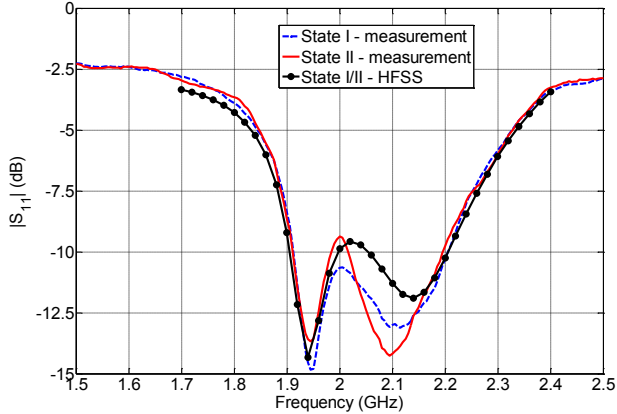

Fig. 7. Simulated and measured reflection coefficients of the antenna prototype for both diodes states.

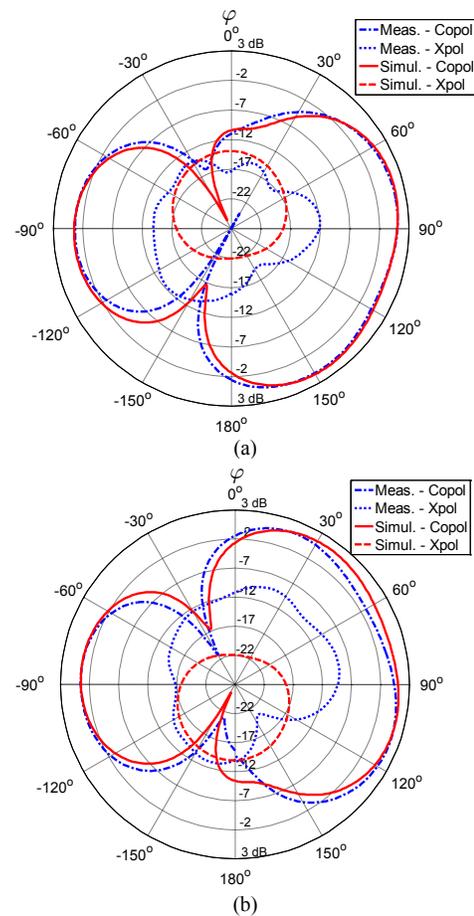

Fig. 8. Simulated and measured instantaneous radiation patterns of the antenna system at 1.95 GHz in the plane of the device platform. (a) $G_1$ (State I). (b) $G_2$ (State II).

TABLE I
RADIATION EFFICIENCY OF THE ANTENNA SYSTEM PROTOTYPE

| Frequency (MHz) | Total Radiation Efficiency | Metallic/Dielectric Losses (dB) | Lumped Elements Losses (dB) |
| --- | --- | --- | --- |
| 1920 | 0.76 | 0.76 | 0.55 |
| 1950 | 0.78 | 0.68 | 0.48 |
| 1980 | 0.83 | 0.52 | 0.33 |
| 2110 | 0.89 | 0.38 | 0.12 |
| 2140 | 0.88 | 0.41 | 0.12 |
| 2170 | 0.87 | 0.45 | 0.11 |

have little contribution to the total loss, namely less than 0.55 dB at all design frequencies.

Based on the obtained instantaneous mirrored patterns, the basis patterns can easily be computed by using (1). The corresponding angular basis patterns of the antenna prototype at 1.95 GHz are shown in Fig. 9. Then, by spatial integration over the full sphere, we can find the power associated with each basis pattern. Fig. 10 shows the distribution of the instantaneous radiated power ($P_{G1}$) across the basis patterns as a function of frequency. The basis has a very good power imbalance ratio of 0.18 dB, 1.25 dB, and 0.09 dB at 1.92 GHz, 1.95 GHz and 1.98 GHz, respectively, namely very close the ideal power imbalance of 0 dB for open-loop MIMO operation. Note that by definition here $P_{G1}$ is normalized to the total input power, so corresponds to the total efficiency of the reconfigurable antenna system (i.e. including mismatch and thermal loss).

Although the used upper-bound criterion in [14] has been shown to be quite efficient for the optimization of beam-space MIMO antennas, it is a rough approximation of real BPSK capacity. In general, the global optimization criteria could be topic of further developments in the emerging field of beam-space MIMO. However, to get some insight into the MIMO performance of the antenna system, the capacity analysis of BPSK signaling based on the obtained simulation results was carried out. We assumed the Kronecker channel model in a narrowband flat-fading operation scenario where two BPSK signals are simultaneously transmitted over two orthogonal basis patterns and then received using a two uncorrelated and uncoupled antenna elements in an open-loop MIMO operation. Fig. 11 shows the capacity of the antenna system for transmit signal to noise ratios (SNRs) of 10 dB and 20 dB versus the operating frequency. The graph also depicts the capacity of an ideal 2×2 classical MIMO system under identical assumptions. The excellent behavior of the designed antenna is confirmed, since the corresponding capacity converges to that of the ideal 2×2 MIMO system.

As stated earlier, the ratio of the second and the first data streams $s_2/s_1$ determines the system state and consequently the states of the p-i-n diodes. Under BPSK signaling, the ratio $s_2/s_1$ remains constant during about 50% of the symbol transitions (i.e. when $s_1$ and $s_2$ do not change or when both change at the same time). In such instants, the states of the p-i-n diodes are not altered and no pattern reconfigurability occurs. Therefore, there is no issue related to the symbol transition in this case, since obviously a pulse shaping filter can be included in the path of the first data stream $s_1$, i.e. the one directly fed to the single RF chain.

However, during other symbol transitions (i.e. when only one of the two symbols changes), the control signal switches the states of the p-i-n diodes and the antenna instantaneous radiation pattern is mirrored. In such symbol transitions, improper transition between the states of the p-i-n diodes may give rise to bandwidth expansion of the transmitted signals. To provide some insight about this out-of-band radiation phenomenon, the spectrum measurements in real propagation conditions were carried out while the antenna system was multiplexing two random BPSK data streams. As shown in Fig. 12, when the antenna system works in a single state (i.e. the case where no p-i-n diode switching occurs), a clean spectral mask is achieved thanks to the pulse shaping filter employed on the RF path. However, when the states of the p-i-n diodes are changed using high slew rate rectangular pulses, the antenna system, as expected, suffers from high out-of-band radiation.

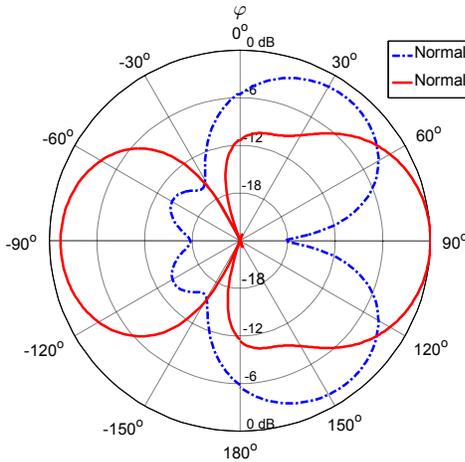

Fig. 9. Simulated magnitude of the angular basis patterns of the antenna system at 1.95 GHz in the plane of the platform. Note that the orthogonality between patterns applies to the complex patterns over the full sphere and not to their magnitudes.

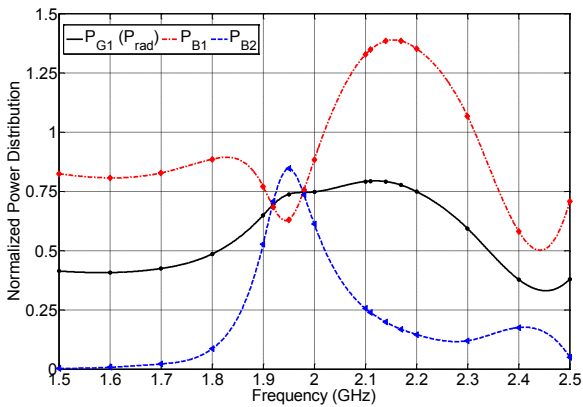

Fig. 10. Normalized power of the instantaneous and basis patterns; $P_{G1}$: instantaneous radiated power, $P_{B1}$: power associated with $B_1$, and $P_{B2}$: power associated with $B_2$. Note that $P_{B1}+P_{B2}=2P_{G1}$ according to (1).

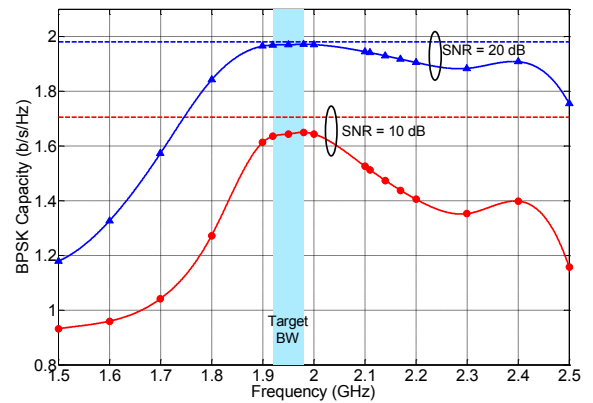

Fig. 11. Capacity under BPSK signaling for two transmit SNR values: 1 (proposed antenna)×2 beam-space MIMO (solid lines) versus 2(ideal antennas)×2 Conventional MIMO (dashed lines).

Though this issue has barely been mentioned before, it is obviously important for beam-space MIMO applications and should be subject of further investigation. One potential solution lies in transient behavior shaping of the p-i-n diodes. A specifically-designed p-i-n diode driver can provide a controlled activation waveform (including reverse bias voltage, forward bias current, timing intervals, etc.) so that the antenna radiation pattern is smoothly reshaped and the transmitted signals properly fit in the allocated frequency band. In a similar manner, the use of varactor diodes as the required reconfigurable components can provide more controllable and smoother impedance changes, but will demand more complex DC biasing networks.

## III. FREQUENCY RECONFIGURABILITY

An important trend in wireless communications is towards a more efficient use of limited spectrum resources. Therefore, and in addition to MIMO techniques, the concept of cognitive radio was brought forward for, among other, achieving dynamic frequency allocation in varying environments [16], [17]. Therefore, in this section we investigate the capability of including frequency reconfiguration in the antenna system design, and thereby extending the low-complexity MIMO benefits for technologies providing dynamic spectrum access such as cognitive radio.

Since the S-parameters and the active port patterns of the three-port radiator are frequency-dependent, the optimization of the antenna yields optimal beam-space MIMO performance around the design frequency only, as can be seen in Fig. 11. Therefore, in order to simultaneously provide frequency reconfigurability and achieve maximum transmission data rate, the instantaneous load values should be selected according to the instant frequency of operation and cannot be simply p-i-n diodes as in the case of fixed frequency operation.

To evaluate the capability of the designed antenna prototype in frequency reconfiguration, we used the S-parameters and the active port radiated fields of the three-port radiator extracted from full-wave simulations to compute the BPSK capacity for a 2D range of reactance $X_1$-$X_2$ under assumptions identical to those in Section II.D. Then, we selected the reactance values which achieve the maximum BPSK capacity at each frequency. Fig. 13 shows that satisfactory capacity is maintained over a bandwidth of 1:2 with a reference of 1.4 b/s/Hz (which corresponds to 83% of the capacity of an ideal 2×2 classical MIMO system for a transmit SNR of 10 dB), when the prototype antenna is properly loaded at each single frequency. As shown in Fig. 14, the use of the optimal reactance values guarantees a return loss better than 10 dB and a power imbalance ratio less than 5 dB on a frequency range larger than 1:2. Fig. 15 plots the optimal reactance values as functions of frequency. Interestingly, plotting a contour map of the BPSK capacity with respect to $X_1$ and $X_2$, such as done in Fig. 16, shows that the performance is in general quite insensitive to moderate variations in the reactance values.

Obviously, the possibility of dynamic frequency operation comes at the cost of more complicated variable load circuits and relies on the availability of appropriate dynamically-controllable reactive loads based on varactor diodes or RF microelectromechanical system (MEMS). A possible method for reducing the complexity of the loads in frequency-flexible operation is to divide the bandwidth into multiple sub-bands, and then to select an optimal reactance pair for each sub-band so that satisfactory beam-space MIMO performance is ensured over the whole bandwidth. In this way, the desired frequency flexibility can be achieved using just a few reactance pairs. To illustrate the applicability of this approach, Fig. 13 shows the BPSK capacity achievable using three different load pairs, which is higher than 1.4 b/s/Hz over a 1:1.5 bandwidth.

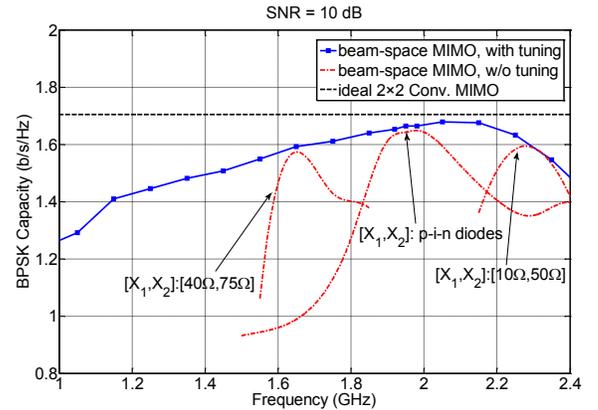

Fig. 13. Maximum achievable capacity under BPSK signaling, when the antenna loadings are properly tuned at each frequency.

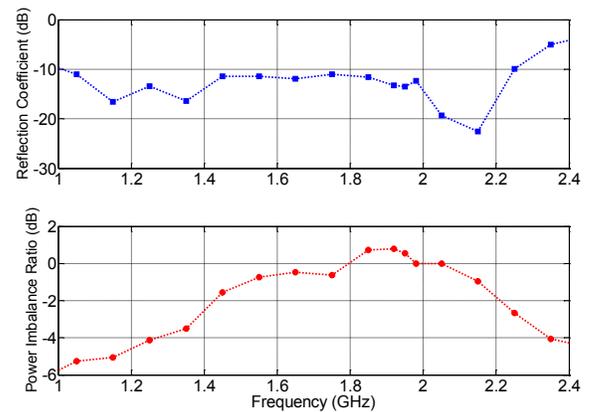

Fig. 14. Corresponding reflection coefficient and power imbalance ratio as functions of frequency, when the antenna loadings are properly tuned at each frequency.

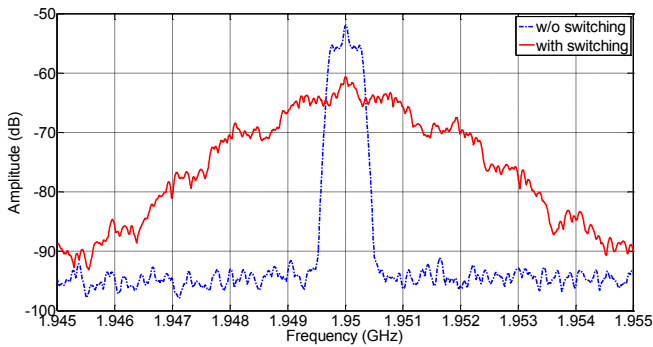

Fig. 12. Out-of-band radiation caused by abrupt switching of the p-i-n diodes when two random BPSK data streams with a symbol rate of 500 Ksymbols/s were being multiplexed by the antenna system. A root-raised cosine pulse shaping filter with a roll-off factor of 0.5 was used on the RF path.

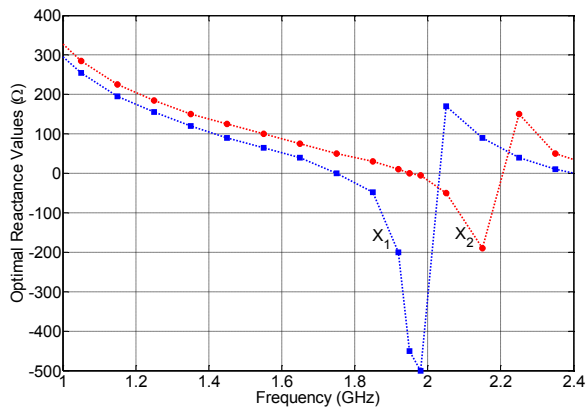

Fig. 15. Required reactance values for achieving maximum BPSK capacity.

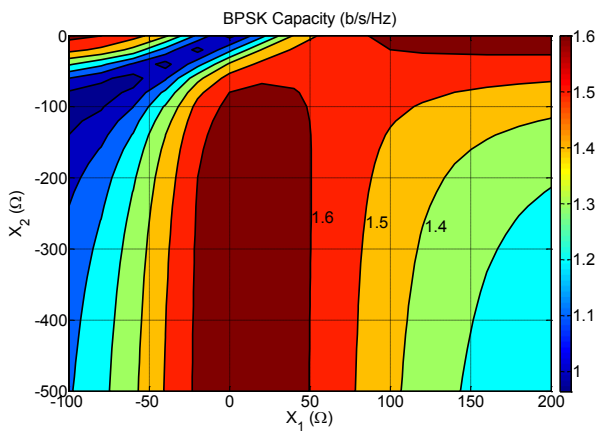

Fig. 16. A contour map of the BPSK capacity with respect to $X_1$ and $X_2$ at 1.95 GHz. The figure shows that a wide range of reactance values can provide an acceptable level of BPSK capacity.

Finally, it is worth mentioning that the developed antenna hardware used here for beam-space MIMO can readily be used to achieve other functionalities. Indeed, as the instantaneous radiation patterns of the antenna are themselves quasi-orthogonal, the hardware can also be employed for receive pattern diversity to dynamically compensate channel variation, or for receive spatial multiplexing by oversampling received signals [18]. Thus, the beam-space MIMO technique can support dynamic selection of the operation mode (transmission/reception) and the operating frequency, paving the way for MIMO-based cognitive radio communications with just a limited increase in hardware complexity.

## IV. CONCLUSION

A compact antenna solution for implementing the beam-space MIMO concept in real small portable devices has been proposed. The solution employs a single-radiator switched antenna, integrated to the device platform, while simplifying the implementation of variable load circuits and DC biasing networks. A practical compact antenna system prototype for single-radio multiplexing of BPSK modulated signals was designed and measured, showing very good agreement with simulations in terms of return loss and radiation patterns. Finally, an early demonstration of the possibility of dynamic frequency allocation in beam-space MIMO transmission was provided. These results constitute significant progress towards the future implementation of the beam-space MIMO concept in real applications.

Future work in this very promising field should be directed towards the issue of the bandwidth expansion associated with transient behavior of the embedded switching elements. On the other hand, modern wireless communications standards demand higher order modulation schemes for higher data rates and lower latency. Therefore, further research needs to be devoted to the design of realistic antenna systems supporting the beam-space multiplexing of higher order modulations. Finally, another future work could address the actual implementation and testing of a frequency-flexible beam-space antenna system.

## ACKNOWLEDGMENT

The authors would like to thank J.-F. Zürcher, P. Belanovic, and A. P. Burg at EPFL for the precious help in prototype fabrication and measurements, and O. N. Alrabadi at Aalborg University for the valuable technical discussions.

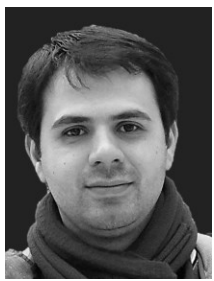

**Mohsen Yousefbeiki** was born in Iran in 1983. He received the B.Sc. and M.Sc. degrees in electrical engineering from University of Tehran, Tehran, Iran, in 2005 and 2008, respectively, and is currently pursuing the Ph.D. degree in electrical engineering at the Ecole Polytechnique Fédérale de Lausanne (EPFL), Lausanne, Switzerland.

His main research interest includes antenna design, particularly for novel radio coding techniques.

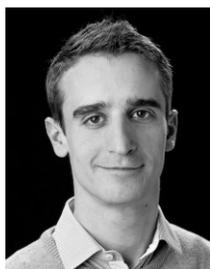

**Julien Perruisseau-Carrier** (S'07-M'09-SM'13) was born in Lausanne, Switzerland, in 1979. He received the M.Sc. and Ph.D. degrees from the Ecole Polytechnique Fédérale de Lausanne (EPFL), Lausanne, Switzerland, in 2003 and 2007, respectively.

In 2003, he was with the University of Birmingham, UK, first as a visiting student and then as a short-term researcher. From 2004 to 2007, he was with the Laboratory of Electromagnetics and Acoustics (LEMA), EPFL, where he completed his PhD while working on various EU funded projects. From 2007 to 2011 he was with the Centre Tecnològic de Telecomunicacions de Catalunya (CTTC), Barcelona, as an associate researcher. Since June 2011 he is a Professor at EPFL funded by the Swiss National Science Foundation, where he leads the group for Adaptive MicroNano Wave Systems. He has led various projects and workpackages at the National, European Space Agency, European Union, and industrial levels. His main research interest concerns interdisciplinary topics related to electromagnetic waves from microwave to terahertz: dynamic reconfiguration, application of micro/nanotechnology, joint antenna-coding techniques, and metamaterials. He has authored +80 and +40 conference and journal papers in these fields, respectively.

Julien Perruisseau-Carrier was the recipient of the Raj Mittra Travel Grant 2010 presented by the IEEE Antennas and Propagation Society, and of the Young Scientist Award of the URSI Intern. Symp. on Electromagnetic Theory, both in 2007 and in 2013. He currently serves as an Associate Editor of the IEEE TRANSACTIONS ON ANTENNAS AND PROPAGATION, as the Swiss representative to URSI's commission B 'Fields and waves', as a member of the Technical committee on RF Nanotechnology (MTT-25) of the IEEE Microwave Theory and Techniques Society, and as the chair of the Working Group on 'Enabling Technologies' of the EU COST Action IC1102.